\documentclass{article}
\begin{document}
{\LARGE
\begin{center}
{\bf
Widths of tetraquarks with open charm}
\end{center}
}

\large

\begin{center}
\vskip3ex
S.M. Gerasyuta $ ^{1,2}$ and V.I. Kochkin $ ^1$

\vskip2ex
$ ^1$ Department of Theoretical Physics, St. Petersburg State University,
198904,

St. Petersburg, Russia

$ ^2$ Department of Physics, LTA, 194021, St. Petersburg, Russia
\end{center}

\vskip2ex

\noindent
E-mail: gerasyuta@SG6488.spb.edu

\vskip4ex
\begin{center}
{\bf Abstract}
\end{center}
\vskip4ex
{\large
In the framework of coupled-channel formalism the relativistic four-quark
equations are found. The dynamical mixing of the meson-meson states
with the four-quark states is considered. The four-quark amplitudes of
the tetraquarks with  open charm, including $u$, $d$, $s$, $c$
quarks, are constructed. The poles of these amplitudes determine the masses
and widths of tetraquarks.


\vskip2ex

\vskip2ex
{\bf I. Introduction.}
\vskip2ex
The discovery by the Belle Collaboration [1] of the very narrow $X(3872)$
resonance has triggered the interest in the charmonium-like states, both
experimentally and theoretically. The observation of $X(3872)$ has been
confirmed by CDF [2], D0 [3] and BaBar Collaboration [4]. Belle
Collaboration observed the $X(3940)$ in double-charmonium production in the
reaction $e^+ e^- \to J/ \psi +X$ [5]. The state, designated as $X(4160)$,
was reported by the Belle Collaboration in Ref. 6. The fact that the newly
found states do not fit quark model calculations [7] has opened the
discussion about the structure of such states. Maiani et al. advocate a
tetraquark explanation for the $X(3872)$ [8, 9]. On the other hand, the
mass of $X(3872)$ is very close to the threshold of $D^*D$ and, therefore,
it can be interpreted as molecular state [10 -- 15]. Several review papers,
as for example [16, 17], discuss the difficulty of interpreting these
resonances as charmonium states.

In the present paper the relativistic four-quark equations are found in the
framework of coupled-channel formalism. The dynamical mixing between the
meson-meson states and the four-quark states is considered [18 -- 20].
Taking the $X(3872)$ and $X(3940)$ as input [21] we predict the masses and
widths of $S$-wave tetraquarks with open charm (Table I).

\vskip2ex
{\bf II. Four-Quark Amplitudes for the Tetraquarks with Open Charm.}
\vskip2ex
We derive the relativistic four-quark equations in the framework of the
dispersion relations technique.

The correct equations for the amplitude are obtained by taking into
account all possible subamplitudes. It corresponds to the division of
complete system into subsystems with the smaller number of particles.
Then one should represent a four-particle amplitude as a sum of six
subamplitudes:

\begin{equation}
A=A_{12}+A_{13}+A_{14}+A_{23}+A_{24}+A_{34}\, . \end{equation}

This defines the division of the diagrams into groups according to the
certain pair interaction of particles. The total amplitude can be
represented graphically as a sum of diagrams.

We need to consider only one group of diagrams and the amplitude
corresponding, for example $A_{12}$. The relativistic generalization of the
Faddeev-Yakubovsky approach [22, 23] for the tetraquark is obtained.
We shall construct the four-quark amplitude of $c \bar u u \bar u$
tetraquark in which the quark amplitudes with quantum numbers of $0^{-+}$
and $1^{--}$ mesons are included. The set of diagrams associated with
the amplitude $A_{12}$ can further be broken down into four groups
corresponding to subamplitudes: $A_1 (s,s_{12},s_{34})$,
$A_2 (s,s_{23},s_{14})$, $A_3 (s,s_{23},s_{123})$, $A_4 (s,s_{14},s_{124})$,
if we consider the tetraquark with the spin-parity $J^{pc}=0^{++}$
($c \bar u u \bar u$).

Here $s_{ik}$ is the two-particle subenergy squared, $s_{ijk}$ corresponds
to the energy squared of particles $i$, $j$, $k$ and $s$ is the system
total energy squared.

In order to represent the subamplitudes  $A_1 (s,s_{12},s_{34})$,
$A_2 (s,s_{23},s_{14})$, $A_3 (s,s_{23},s_{123})$,
$A_4 (s,s_{14},s_{124})$  in the form of dispersion relations it is
necessary to define the amplitudes of quark-antiquark interaction
$a_n(s_{ik})$. The pair quarks amplitudes $q \bar q\rightarrow q \bar q$
are calculated in the framework of the dispersion $N/D$ method with the
input four-fermion interaction [24 -- 26] and the quantum numbers of the
gluon [27]. The regularization of the dispersion integral for the
$D$-function is carried out with the cutoff parameter $\Lambda$.
The four-quark interaction is considered as an input [27]:

\begin{equation}
g_V \left(\bar q \lambda I_f \gamma_{\mu} q \right)^2 +
g^{(s)}_V \left(\bar q \lambda I_f \gamma_{\mu} q \right)
\left(\bar s \lambda \gamma_{\mu} s \right)+
g^{(ss)}_V \left(\bar s \lambda \gamma_{\mu} s \right)^2 \, . \end{equation}

\noindent
Here $I_f$ is the unity matrix in the flavor space $(u, d)$. $\lambda$ are
the color Gell-Mann matrices. Dimensional constants of the four-fermion
interaction $g_V$, $g^{(s)}_V$ and $g^{(ss)}_V$ are parameters of the
model. At $g_V =g^{(s)}_V =g^{(ss)}_V$ the flavor $SU(3)_f$ symmetry occurs.
The strange quark violates the flavor $SU(3)_f$ symmetry. In order to avoid
an additional violation parameters, we introduce the scale shift of the
dimensional parameters [27]:

\begin{equation}
g=\frac{m^2}{\pi^2}g_V =\frac{(m+m_s)^2}{4\pi^2}g_V^{(s)} =
\frac{m_s^2}{\pi^2}g_V^{(ss)} \, .\end{equation}

\begin{equation}
\Lambda=\frac{4\Lambda(ik)}
{(m_i+m_k)^2}. \end{equation}

\noindent
Here $m_i$ and $m_k$ are the quark masses in the intermediate state of
the quark loop. Dimensionless parameters $g$ and $\Lambda$ are supposed
to be constants which are independent of the quark interaction type. The
applicability of Eq. (2) is verified by the success of
De Rujula-Georgi-Glashow quark model [28], where only the short-range
part of Breit potential connected with the gluon exchange is
responsible for the mass splitting in hadron multiplets.

We use the results of our relativistic quark model [27] and write down
the pair quarks amplitude in the form:

\begin{equation}
a_n(s_{ik})=\frac{G^2_n(s_{ik})}
{1-B_n(s_{ik})} \, ,\end{equation}

\begin{equation}
B_n(s_{ik})=\int\limits_{(m_i+m_k)^2}^{\frac{(m_i+m_k)^2\Lambda}{4}}
\hskip2mm \frac{ds'_{ik}}{\pi}\frac{\rho_n(s'_{ik})G^2_n(s'_{ik})}
{s'_{ik}-s_{ik}} \, .\end{equation}

\noindent
Here $G_n(s_{ik})$ are the quark-antiquark vertex functions. The vertex
functions are determined by the contribution of the crossing channels.
The vertex functions satisfy the Fierz relations. All of these vertex
functions are generated from $g_V$, $g^{(s)}_V$ and $g^{(ss)}_V$.
$B_n(s_{ik})$, $\rho_n (s_{ik})$ are the Chew-Mandelstam functions with
cutoff $\Lambda$ and the phase spaces, respectively.

Here $n=1$ determines a $q \bar q$-pairs with $J^{pc}=0^{-+}$ in the $1_c$
color state, $n=2$ corresponds to a $q \bar q$-pairs with $J^{pc}=1^{--}$
in the $1_c$ color state, and $n=3$ defines the $q \bar q$-pairs
corresponding to tetraquarks with quantum numbers: $J^{pc}=0^{++}$,
$1^{++}$, $2^{++}$.

In the case in question, the interacting quarks do not produce a bound
state; therefore, the integration in Eqs. (7) -- (10) is carried out from
the threshold $(m_i+m_k)^2$ to the cutoff $\Lambda(i,k)$.
The coupled integral equation systems (the tetraquark state with $n=3$
and $J^{pc}=0^{++}$ for the $c \bar u u \bar u$) can be described as:

\begin{eqnarray}
A_1(s,s_{12},s_{34})&=&\frac{\lambda_1 B_2(s_{12})  B_2(s_{34})}
{[1- B_2(s_{12})][1- B_2(s_{34})]}+
2\hat J_2(s_{12},s_{34},2,2) A_3(s,s'_{23},s'_{123})\nonumber\\
&&\nonumber\\
&+&2\hat J_2(s_{12},s_{34},2,2) A_4(s,s'_{14},s'_{124}) \, ,\\
&&\nonumber\\
A_2(s,s_{23},s_{14})&=&\frac{\lambda_2 B_1(s_{23})  B_1(s_{14})}
{[1- B_1(s_{23})][1- B_1(s_{14})]}+
2\hat J_2(s_{23},s_{14},1,1) A_3(s,s'_{34},s'_{234})\nonumber\\
&&\nonumber\\
&+&2\hat J_2(s_{23},s_{14},1,1) A_4(s,s'_{12},s'_{123}) \, ,\\
&&\nonumber\\
A_3(s,s_{23},s_{123})&=&\frac{\lambda_3 B_3(s_{23})}{[1- B_3(s_{23})]}+
2\hat J_3(s_{23},3) A_1(s,s'_{12},s'_{34})
+\hat J_3(s_{23},3) A_2(s,s'_{12},s'_{34})\nonumber\\
&&\nonumber\\
&+&\hat J_1(s_{23},3) A_4(s,s'_{34},s'_{234})+
\hat J_1(s_{23},3) A_3(s,s'_{12},s'_{123}) \, ,\\
&&\nonumber\\
A_4(s,s_{14},s_{124})&=&\frac{\lambda_4 B_3(s_{14})}{[1- B_3(s_{14})]}+
2\hat J_3(s_{14},3) A_1(s,s'_{13},s'_{24})
+2\hat J_3(s_{14},3) A_2(s,s'_{13},s'_{24})\nonumber\\
&&\nonumber\\
&+&2\hat J_1(s_{14},3) A_3(s,s'_{14},s'_{134})
+2\hat J_1(s_{14},3) A_4(s,s'_{14},s'_{134}) \, ,
\end{eqnarray}

\noindent
where $\lambda_i$, $i=1, 2, 3, 4$ are the current constants. They do not
affect the mass spectrum of tetraquarks. We introduce the integral
operators:

\begin{eqnarray}
\hat J_1(s_{12},l)&=&\frac{G_l(s_{12})}
{[1- B_l(s_{12})]} \int\limits_{(m_1+m_2)^2}^{\frac{(m_1+m_2)^2\Lambda}{4}}
\frac{ds'_{12}}{\pi}\frac{G_l(s'_{12})\rho_l(s'_{12})}
{s'_{12}-s_{12}} \int\limits_{-1}^{+1} \frac{dz_1}{2} \, ,\\
&&\nonumber\\
\hat J_2(s_{12},s_{34},l,p)&=&\frac{G_l(s_{12})G_p(s_{34})}
{[1- B_l(s_{12})][1- B_p(s_{34})]}
\int\limits_{(m_1+m_2)^2}^{\frac{(m_1+m_2)^2\Lambda}{4}}
\frac{ds'_{12}}{\pi}\frac{G_l(s'_{12})\rho_l(s'_{12})}
{s'_{12}-s_{12}}\nonumber\\
&&\nonumber\\
&\times&\int\limits_{(m_3+m_4)^2}^{\frac{(m_3+m_4)^2\Lambda}{4}}
\frac{ds'_{34}}{\pi}\frac{G_p(s'_{34})\rho_p(s'_{34})}
{s'_{34}-s_{34}}
\int\limits_{-1}^{+1} \frac{dz_3}{2} \int\limits_{-1}^{+1} \frac{dz_4}{2}
 \, ,\\
&&\nonumber\\
\hat J_3(s_{12},l)&=&\frac{G_l(s_{12},\tilde \Lambda)}
{[1- B_l(s_{12},\tilde \Lambda)]} \, \, \frac{1}{4\pi}
\int\limits_{(m_1+m_2)^2}^{\frac{(m_1+m_2)^2\tilde \Lambda}{4}}
\frac{ds'_{12}}{\pi}\frac{G_l(s'_{12},\tilde \Lambda)
\rho_l(s'_{12})}
{s'_{12}-s_{12}}\nonumber\\
&&\nonumber\\
&\times&\int\limits_{-1}^{+1}\frac{dz_1}{2}
\int\limits_{-1}^{+1} dz \int\limits_{z_2^-}^{z_2^+} dz_2
\frac{1}{\sqrt{1-z^2-z_1^2-z_2^2+2zz_1z_2}} \, ,
\end{eqnarray}

\noindent
here $l$, $p$ are equal to $1-3$.

In Eqs. (11) and (13) $z_1$ is the cosine of the angle between the relative
momentum of the particles 1 and 2 in the intermediate state and the momentum
of the particle 3 in the final state, taken in the c.m. of particles
1 and 2. In Eq. (13) $z$ is the cosine of the angle between the momenta
of particles 3 and 4 in the final state, taken in the c.m. of particles
1 and 2. $z_2$ is the cosine of the angle between the relative
momentum of particles 1 and 2 in the intermediate state and the momentum
of the particle 4 in the final state, is taken in the c.m. of particles
1 and 2. In Eq. (10): $z_3$ is the cosine of the angle between relative
momentum of particles 1 and 2 in the intermediate state and the relative
momentum of particles 3 and 4 in the intermediate state, taken in the c.m.
of particles 1 and 2. $z_4$ is the cosine of the angle between the relative
momentum of the particles 3 and 4 in the intermediate state and that of the
momentum of the particle 1 in the intermediate state, taken in the c.m.
of particles 3, 4.

We can pass from the integration over the cosines of the angles to the
integration over the subenergies [29].

Let us extract two-particle singularities in the amplitudes
$A_1(s,s_{12},s_{34})$, $A_2 (s,s_{23},s_{14})$, $A_3 (s,s_{23},s_{123})$
and $A_4 (s,s_{14},s_{124})$:

\begin{equation}
A_1(s,s_{ik},s_{lm})=\frac{\alpha_1(s,s_{ik},s_{lm})B_2(s_{ik})B_2(s_{lm})}
{[1-B_2(s_{ik})][1-B_2(s_{lm})]} \, ,\end{equation}

\begin{equation}
A_2(s,s_{ik},s_{lm})=\frac{\alpha_2(s,s_{ik},s_{lm})B_1(s_{ik})B_1(s_{lm})}
{[1-B_1(s_{ik})][1-B_1(s_{lm})]} \, ,\end{equation}

\begin{equation}
A_j(s,s_{ik},s_{ikl})=\frac{\alpha_j(s,s_{ik},s_{ikl})B_3(s_{ik})}
{1-B_3(s_{ik})} \, , \quad\quad j=3-4 \, .\end{equation}

We do not extract three-particles singularities, because they are weaker
than two-particle singularities.

We used the classification of singularities, which was proposed in
paper [30]. The construction of the approximate solution of Eqs.
(7) -- (10) is based on the extraction of the leading singularities
of the amplitudes. The main singularities in $s_{ik}\approx (m_i+m_k)^2$
are from pair rescattering of the particles $i$ and $k$. First of all there
are threshold square-root singularities. Also possible are pole
singularities which correspond to the bound states. The amplitudes
apart from two-particle singularities have triangular singularities and the
singularities defining the interactions of four particles. Such
classification allows us to search the corresponding solution of Eqs.
(7) -- (10) by taking into account some definite number of leading
singularities and neglecting all the weaker ones. We consider the
approximation which defines two-particle, triangle and four-particle
singularities. The functions $\alpha_1(s,s_{12},s_{34})$,
$\alpha_2(s,s_{23},s_{14})$, $\alpha_3(s,s_{23},s_{123})$ and
$\alpha_4(s,s_{14},s_{124})$ are the smooth functions of $s_{ik}$,
$s_{ikl}$, $s$ as compared with the singular part of the amplitude, hence
they can be expanded in a series in the singulary point and only the first
term of this series should be employed further. Using this classification,
one defines the reduced amplitudes $\alpha_1$, $\alpha_2$, $\alpha_3$,
$\alpha_4$ as well as the $B$-functions in the middle point of physical
region of Dalitz-plot at the point $s_0$:

\begin{eqnarray}
s_0^{ik}=0.25(m_i+m_k)^2 s_0 \, ,\nonumber
\end{eqnarray}

\begin{equation}
s_{123}=0.25 s_0 \sum\limits_{i,k=1 \atop i\ne k}^{3} (m_i+m_k)^2 -
\sum\limits_{i=1}^{3} m_i^2 \, ,
\quad s_0=\frac{s+2\sum\limits_{i=1}^{4} m_i^2}
{0.25 \sum\limits_{i,k=1 \atop i\ne k}^{4} (m_i+m_k)^2}
\, . \end{equation}

Such a choice of point $s_0$ allows us to replace integral Eqs. (7) -- (10)
by the algebraic equations (18) -- (21) respectively:

\begin{equation}
\alpha_1=\lambda_1+2\alpha_3 JB_1 (2,2,3)+2\alpha_4 JB_2 (2,2,3)
\, ,\end{equation}

\begin{equation}
\alpha_2=\lambda_2+2\alpha_3 JB_3 (1,1,3)+2\alpha_4 JB_4 (1,1,3)
\, , \end{equation}

\begin{equation}
\alpha_3=\lambda_3+2\alpha_1 JC_1 (3,2,2)+2\alpha_2 JC_2 (3,1,1)
+\alpha_4 JA_1 (3)+\alpha_3 JA_2 (3)\, , \end{equation}

\begin{equation}
\alpha_4=\lambda_4+2\alpha_1 JC_3 (3,2,2)+2\alpha_2 JC_4 (3,1,1)
+\alpha_3 JA_3 (3)+\alpha_4 JA_4 (3)\, . \end{equation}

We use the functions $JA_i(l)$, $JB_i(l,p,r)$, $JC_i(l,p,r)$
$(l,p,r=1-3)$, which are determined by the various $s_0^{ik}$ (Eq. 17).
These functions are similar to the functions:

\begin{eqnarray}
JA_4(l)&=&\frac{G_l^2(s_0^{12})B_l^2(s_0^{23})}
{B_l(s_0^{12})} \int\limits_{(m_1+m_2)^2}^{\frac{(m_1+m_2)^2\Lambda}{4}}
\frac{ds'_{12}}{\pi}\frac{\rho_l(s'_{12})}
{s'_{12}-s_{12}} \int\limits_{-1}^{+1} \frac{dz_1}{2}
\frac{1}{1-B_l (s'_{23})} \, ,\\
&&\nonumber\\
JB_1(l,p,r)&=&\frac{G_l^2(s_0^{12})G_p^2(s_0^{34})B_r(s_0^{23})}
{B_l(s_0^{12})B_p(s_0^{34})}
\int\limits_{(m_1+m_2)^2}^{\frac{(m_1+m_2)^2\Lambda}{4}}
\frac{ds'_{12}}{\pi}\frac{\rho_l(s'_{12})}
{s'_{12}-s_{12}}\nonumber\\
&&\nonumber\\
&\times&\int\limits_{(m_3+m_4)^2}^{\frac{(m_3+m_4)^2\Lambda}{4}}
\frac{ds'_{34}}{\pi}\frac{\rho_p(s'_{34})}{s'_{34}-s_{34}}
\int\limits_{-1}^{+1} \frac{dz_3}{2} \int\limits_{-1}^{+1} \frac{dz_4}{2}
\frac{1}{1-B_r (s'_{23})} \, ,\\
&&\nonumber\\
JC_3(l,p,r)&=&\frac{G_l^2(s_0^{12},\tilde \Lambda)B_p(s_0^{23})
B_r(s_0^{14})}{1- B_l(s_0^{12},\tilde \Lambda)}
\frac{1-B_l(s_0^{12})}{B_l(s_0^{12})}\, \, \frac{1}{4\pi}
\int\limits_{(m_1+m_2)^2}^{\frac{(m_1+m_2)^2\tilde \Lambda}{4}}
\frac{ds'_{12}}{\pi}\frac{\rho_l(s'_{12})}{s'_{12}-s_{12}}\nonumber\\
&&\nonumber\\
&\times&\int\limits_{-1}^{+1}\frac{dz_1}{2}
\int\limits_{-1}^{+1} dz \int\limits_{z_2^-}^{z_2^+} dz_2
\frac{1}{\sqrt{1-z^2-z_1^2-z_2^2+2zz_1z_2}}\nonumber\\
&&\nonumber\\
&\times&\frac{1}{[1-B_p(s'_{23})][1-B_r(s'_{14})]} \, ,
\end{eqnarray}

\begin{eqnarray}
\tilde \Lambda(ik)=\left\{ \Lambda(ik), \hskip5.7em  {\rm if} \quad
\Lambda(ik) \le (\sqrt{s_{123}}+m_3)^2 \atop
(\sqrt{s_{123}}+m_3)^2, \hskip2em {\rm if} \quad
\Lambda(ik) > (\sqrt{s_{123}}+m_3)^2 \right.
\end{eqnarray}

The other choices of point $s_0$ do not change essentially the contributions
of $\alpha_1$, $\alpha_2$, $\alpha_3$ and $\alpha_4$, therefore we omit
the indices $s_0^{ik}$. Since the vertex functions depend only slightly
on energy it is possible to treat them as constants in our approximation.

The solutions of the system of equations are considered as:

\begin{equation}
\alpha_i(s)=F_i(s,\lambda_i)/D(s) \, ,\end{equation}

\noindent
where zeros of $D(s)$ determinants define the masses of bound states of
tetraquarks. $F_i(s,\lambda_i)$ determine the contributions of
subamplitudes to the tetraquark amplitude.

\vskip2ex
{\bf III. Calculation results.}
\vskip2ex
Our calculations do not include the new parameters. We use the cutoff
$\Lambda=10.0$ and the gluon coupling constant $g=0.794$, which are
determined by fixing the tetraquark masses for the states  with the hidden
charm $J^{pc}=1^{++}$ $X(3872)$ and $J^{pc}=2^{++}$ $X(3940)$ [21].
The quark masses of model $m_{u,d}=385\, MeV$ and $m_s=510\, MeV$ coincide
with the ordinary meson model ones [27].
In order to fix anyhow $m_c=1586\, MeV$, we use the tetraquark mass for
the $J^{pc}=2^{++}$ $X(3940)$. The masses and widths of meson-meson states
with the spin-parity $J^{pc}=0^{++}$, $1^{++}$, $2^{++}$ are given in
Table I. In our paper we predicted the scalar tetraquark with the mass
$M=2610\, MeV$ and the width $\Gamma_{0^{++}}=180\, MeV$
(channel $D^0 \eta$). We calculated the scalar tetraquark
with the mass $M=2691\, MeV$ and the width $\Gamma_{0^{++}}=110\, MeV$
(channels $D_s^+ \eta$ and $D^0 K^+$).
The other scalar tetraquark is predicted as $M=2805\, MeV$
and the width $\Gamma_{0^{++}}=80\, MeV$ (channel $D^+_s K^-$). The
tetraquarks with the spin-parity $J^{pc}=1^{++}$, $2^{++}$ (Table I) have
only the weak decays.

The functions $F_i(s,\lambda_i)$ (Eq. (26)) allow us to obtain the overlap
factors $f$ for the tetraquarks. We calculated the overlap factors $f$
and the phase spaces $\rho$ for the reactions $X\to M_1 M_2$ (Table II).
The widths of the scalar tetraquarks with open charm are obtained (Table I).
We considered the formula $\Gamma \sim f^2 \times \rho$ [31], there $\rho$
is the phase space. The widths of the tetraquarks are fitted by the fixing
width $\Gamma_{2^{++}}=(39\pm 26) \, MeV$ [21] for the tetraquark $X(3940)$
$(c \bar c u \bar u)$ with the spin-parity $J^{pc}=2^{++}$.

In the open charm sector the scalar tetraquarks have relatively small
width $\sim 100-200 \, MeV$, so in principle these exotic states could be
observed. The low-lying tetraquarks with the open charm were calculated
in other works [32, 33]. These states appear as narrow states. In our
model the tetraquarks with the open charm and the spin-parity
$J^{pc}=1^{++}$, $2^{++}$ can decay only in the weak channels.

\vskip2.0ex
{\bf Acknowledgments.}
\vskip2.0ex

The work was carried with the support of the Russian Ministry of Education
(grant 2.1.1.68.26).

\newpage

\noindent
Table I. Masses and widths of tetraquark with open charm.

\vskip1.5ex
\noindent
Parameters of model [21]: quark masses $m_{u,d}=385\, MeV$, $m_s=510\, MeV$
and $m_c=1586\, MeV$; cutoff parameter $\Lambda=10.0$, gluon coupling
constant $g=0.794$.

\vskip1.5ex

\noindent
\begin{tabular}{|c|c|c|c|c|c|c|c|}
\hline
Tetraquark & $J^{pc}$ & Mass ($MeV$) & Width ($MeV$)
 & $J^{pc}$ & Mass ($MeV$) & $J^{pc}$ & Mass ($MeV$) \\[5pt]
\hline
($c \bar u$)($u \bar u$) & $0^{++}$ & $2610$ & $180$
& $1^{++}$ & $2672$ & $2^{++}$ & $2736$ \\[3pt]
\hline
\begin{tabular}{c}
($c \bar s$)($u \bar u$)\\
($c \bar u$)($u \bar s$)
\end{tabular}
& $0^{++}$ & $2691$ & $110$
& $1^{++}$ & $2770$ & $2^{++}$ & $2851$ \\
\hline
\begin{tabular}{c}
($c \bar u$)($s \bar s$)\\
($c \bar s$)($s \bar u$)
\end{tabular}
& $0^{++}$ & $2805$ & $80$
& $1^{++}$ & $2890$ & $2^{++}$ & $2975$ \\
\hline
\end{tabular}

\vskip12ex

\noindent
Table II. Overlap factors $f$ and phase spaces $\rho$ of  tetraquarks
with open charm.

\vskip1.5ex

\noindent
\begin{tabular}{|cc|c|c|c|}
\hline
Tetraquark & (channels) & $J^{pc}$ & $f$ & $\rho$ \\[5pt]
\hline
($c \bar u$)($u \bar u$) & \hskip-0.8em $D^0 \eta$ & $0^{++}$ & $0.396$ &
$0.325$ \\[3pt]
\hline
\begin{tabular}{c}
($c \bar s$)($u \bar u$)\\
($c \bar u$)($u \bar s$)
\end{tabular}
&
\begin{tabular}{l}
$D^+_s \eta$ \\
$D^0 K^+$
\end{tabular}
& $0^{++}$ &
\begin{tabular}{c}
$0.246$ \\
$0.183$
\end{tabular}
&
\begin{tabular}{c}
$0.300$ \\
$0.414$
\end{tabular}
\\
\hline
\begin{tabular}{c}
($c \bar u$)($s \bar s$)\\
($c \bar s$)($s \bar u$)
\end{tabular}
&
\begin{tabular}{l}
$D^0 \eta_s$ \\
$D^+_s K^-$
\end{tabular}
& $0^{++}$ &
\begin{tabular}{c}
$0.192$ \\
$0.237$
\end{tabular}
&
\begin{tabular}{c}
-- \\
$0.407$
\end{tabular}
\\
\hline
\end{tabular}

\newpage
{\bf \Large References.}
\vskip5ex
\noindent
1. S.K. Choi et al. (Belle Collaboration), Phys. Rev. Lett. {\bf 91},
262001 (2003).

\noindent
2. D. Acosta et al. (CDF Collaboration), Phys. Rev. Lett. {\bf 93},
072001 (2004).

\noindent
3. V.M. Abazov et al. (D0 Collaboration), Phys. Rev. Lett. {\bf 93},
162002 (2004).

\noindent
4. B. Aubert et al. (BaBar Collaboration), Phys. Rev. D{\bf 71},
071103 (2005).

\noindent
5. K. Abe et al. (Belle Collaboration), Phys. Rev. Lett. {\bf 98},
082001 (2007).

\noindent
6. I. Adachi et al. (Belle Collaboration), arXiv: 0708.3812 [hep-ex].

\noindent
7. S. Godfrey and N. Isgur, Phys. Rev. D{\bf 32}, 189 (1985).

\noindent
8. L. Maiani, F. Piccinini, A.D. Polosa and V. Riequer,
Phys. Rev. D{\bf 71}, 014028

(2005).

\noindent
9. L. Maiani, A.D. Polosa and V. Riequer, Phys. Rev. Lett. {\bf 99},
182003 (2007).

\noindent
10. N.A. Tornqvist, Phys. Lett. B{\bf 590}, 209 (2004).

\noindent
11. F.E. Close and P.R. Page, Phys. Lett. B{\bf 628}, 215 (2005).

\noindent
12. E.S. Swanson, Int. J. Mod. Phys. A{\bf 21}, 733 (2006).

\noindent
13. T. Barnes, Int. J. Mod. Phys. A{\bf 21}, 5583 (2006).

\noindent
14. S.H. Lee, K. Morita and M. Nielsen, arXiv: 0808.3168 [hep-ph].

\noindent
15. Y. Dong, A. Faessler, T. Gutsche and V.E. Lyubovitskij,
arXiv: 0802.3610 [hep-ph].

\noindent
16. E.S. Swanson, Phys. Rept. {\bf 429}, 243 (2006).

\noindent
17. S. Godfrey and S.L. Olsen, arXiv: 0801.3867 [hep-ph].

\noindent
18. S.M. Gerasyuta and V.I. Kochkin, Z. Phys. C{\bf 74}, 325 (1997).

\noindent
19. S.M. Gerasyuta and V.I. Kochkin, Nuovo Cim. A{\bf 110}, 1313 (1997).

\noindent
20. S.M. Gerasyuta and V.I. Kochkin, arXiv: 0804.4567 [hep-ph].

\noindent
21. S.M. Gerasyuta and V.I. Kochkin, arXiv: 0809.1758 [hep-ph].

\noindent
22. O.A. Yakubovsky, Sov. J. Nucl. Phys. {\bf 5}, 1312 (1967).

\noindent
23. S.P. Merkuriev and L.D. Faddeev, Quantum scattering theory for system
of few

particles (Nauka, Moscow 1985) p. 398.

\noindent
24. Y. Nambu and G. Jona-Lasinio, Phys. Rev. {\bf 122}, 365 (1961):
ibid. {\bf 124}, 246

(1961).

\noindent
25. T. Appelqvist and J.D. Bjorken, Phys. Rev. D{\bf 4}, 3726 (1971).

\noindent
26. C.C. Chiang, C.B. Chiu, E.C.G. Sudarshan and X. Tata,
Phys. Rev. D{\bf 25}, 1136

(1982).

\noindent
27. V.V. Anisovich, S.M. Gerasyuta and A.V. Sarantsev,
Int. J. Mod. Phys. A{\bf 6}, 625

(1991).

\noindent
28. A.De Rujula, H.Georgi and S.L.Glashow, Phys. Rev. D{\bf 12}, 147 (1975).

\noindent
29. S.M. Gerasyuta and V.I. Kochkin, Yad. Fiz. {\bf 59}, 512 (1996)
[Phys. At. Nucl. {\bf 59},

484 (1996)].

\noindent
30. V.V. Anisovich and A.A. Anselm, Usp. Phys. Nauk. {\bf 88}, 287 (1966).

\noindent
31. J.J. Dudek and F.E. Close, Phys. Lett. B{\bf 583}, 278 (2004).

\noindent
32. E.E. Kolomeitsev and M.F.M. Lutz, Phys. Lett. B{\bf 582}, 39 (2004).

\noindent
33. F.K. Guo, P.N. Shen and H.C. Chiang, Phys. Lett. B{\bf 647}, 133
(2007).

\end{document}